\documentclass[%
reprint,
superscriptaddress,
amsmath,amssymb,
]{revtex4-2}

\usepackage{graphicx}% Include figure files
\usepackage{bm}% bm=bold math
\usepackage[hidelinks]{hyperref}

\begin{document}

\title{Stabilization of sliding ferroelectricity through exciton condensation
}

\author{Matteo D'Alessio}
\affiliation{University of Modena e Reggio Emilia, Department of Physics, Informatics and Mathematics, Via Campi 213a, 41125 Modena, Italy}
\affiliation{CNR---Istituto Nanoscienze, Via Campi 213a, 41125 Modena, Italy}
\author{Daniele Varsano}
\affiliation{CNR---Istituto Nanoscienze, Via Campi 213a, 41125 Modena, Italy}
\author{Elisa Molinari}
\affiliation{University of Modena e Reggio Emilia, Department of Physics, Informatics and Mathematics, Via Campi 213a, 41125 Modena, Italy}
\affiliation{CNR---Istituto Nanoscienze, Via Campi 213a, 41125 Modena, Italy}
\author{Massimo Rontani*}
\affiliation{CNR---Istituto Nanoscienze, Via Campi 213a, 41125 Modena, Italy}

%\date{\today}

\maketitle

%\tableofcontents

%%%%%%%%%%% PAPER STARTS HERE

%%%%%%%%%%%%%%%%%%%%%%%%%%%%%%%%%%%%%%%%%%%%%%
%%%%%%%%%%%%%%%%%% ABSTRACT %%%%%%%%%%%%%%%%%%
\noindent Sliding ferroelectricity is a phenomenon that arises from the insurgence of spontaneous electronic polarization perpendicular to the layers of two-dimensional (2D) systems upon the relative sliding of the atomic layer constituents. 
Because of the weak van der Waals (vdW) interactions between layers, sliding and the associated symmetry breaking can occur at low energy cost in materials such as transition-metal dichalcogenides. 
Here we discuss theoretically the origin and quantitative understanding of the phenomenon by focusing on a prototype structure, the WTe$_2$ bilayer, where sliding ferroelectricity was first experimentally observed. 
We show that excitonic effects induce relevant energy band renormalizations in the ground state, and exciton condensation contributes significantly to stabilizing ferroelectricity upon sliding, beyond previous predictions that disregard electron-hole interaction effects. 
Enhanced excitonic effects in 2D and vdW sliding are general phenomena that point to sliding ferroelectricity as relevant for a broad class of important materials, where the intrinsic electric dipole can couple with other quantum phenomena and, in turn, an external electric field can control the quantum phases through ferroelectricity in unexplored ways. 
% 173 WORDS  

%%%%%%%%%%%%%%%%%%%%%%%%%%%%%%%%%%%%%%%%%%%%%%
%%%%%%%%%%%%%%% INTRODUCTION %%%%%%%%%%%%%%%%%
\vspace{0.2cm}
\noindent\textbf{Introduction} \\
Since the discovery of ferroelectricity in the layered semimetal WTe$_2$ \cite{fei_ferroelectric_2018} and the proposed sliding mechanism \cite{yang_origin_2018}, the interest in sliding ferroelectricity has boomed and the pool of 2D candidate materials has significantly grown \cite{wu_sliding_2021, wang_towards_2023, de_la_barrera_direct_2021, vizner_stern_interfacial_2021, ma_tunable_2021, zhang_ferroelastic_2021, niu_giant_2022, meng_sliding_2022, miao_direct_2022, wang_interfacial_2022, sui_sliding_2023,  wan_room_temperature_2022, liu_tunable_2023, wang_sliding_2023, yang_across-layer_2023, ding_quasi_2024, pakdel_high-throughput_2024, zhang_electronic_2024, wei_in-plane_2024}. 
This interest is motivated by both fundamental scientific questions and potential technological applications.
Interlayer sliding is an intrinsic mechanism, and it can  drive ferroelectricity in an important class of materials, where it can couple to other quantum phenomena of great scientific interest \cite{park_ferroelectric_2019, ding_phase_2021, liang_intertwined_2021, xiao_non-synchronous_2022, jindal_coupled_2023, shi_revealing_2024, zhang_ferroelectric_2024}. 
Reciprocally, through ferroelectricity an external vertical electric field could control quantum phases of these 2D systems in unexplored ways, including e.g.~topological, ferromagnetic or optical transitions \cite{xiao_berry_2020, liu_magnetoelectric_2020, zhou_photo-magnetization_2022, chen_enhanced_2023, gao_large_2024, song_coexistence_2024, liang_sliding_2025, niu_ferroelectricity_2025, yang_sliding_2025, sevik_state_2025, zhu_sliding_2025}. 
From the technological point of view, the search for efficient 2D ferroelectrics is considerably more promising in vdW materials than in conventional semiconductors: Owing to the weak vdW interaction between atomic planes, combined with strong in-plane bonding, sliding parallel to the 2D planes takes place with limited energy cost and without significant vertical distortions \cite{reguzzoni2012potential, Levita2014sliding}. 
This could, for example, enable high-speed data writing and memory devices, as well as the integration of multiple functions in future nanoelectronics and spintronics \cite{bian_developing_2024, chen_strong_2024, yasuda_ultrafast_2024, liang_nanosecond_2025, bian_high-performance_2023}. 

A full quantitative understanding of sliding ferroelectricity is however still lacking. 
To illustrate this point, let us focus on the paradigmatic case of the WTe$_2$ bilayer, the first system where it was observed experimentally and attributed to interlayer charge redistribution, in spite of its metallic character \cite{fei_ferroelectric_2018}. 
Bilayer WTe$_2$ is attractive because it was experimentally shown that its polarization does not vanish up to room temperature; 
moreover, it retains its switching capability at 300 K even when embedded in 2D devices \cite{fei_ferroelectric_2018}.

\begin{figure*}
\includegraphics[width=1.0\linewidth]{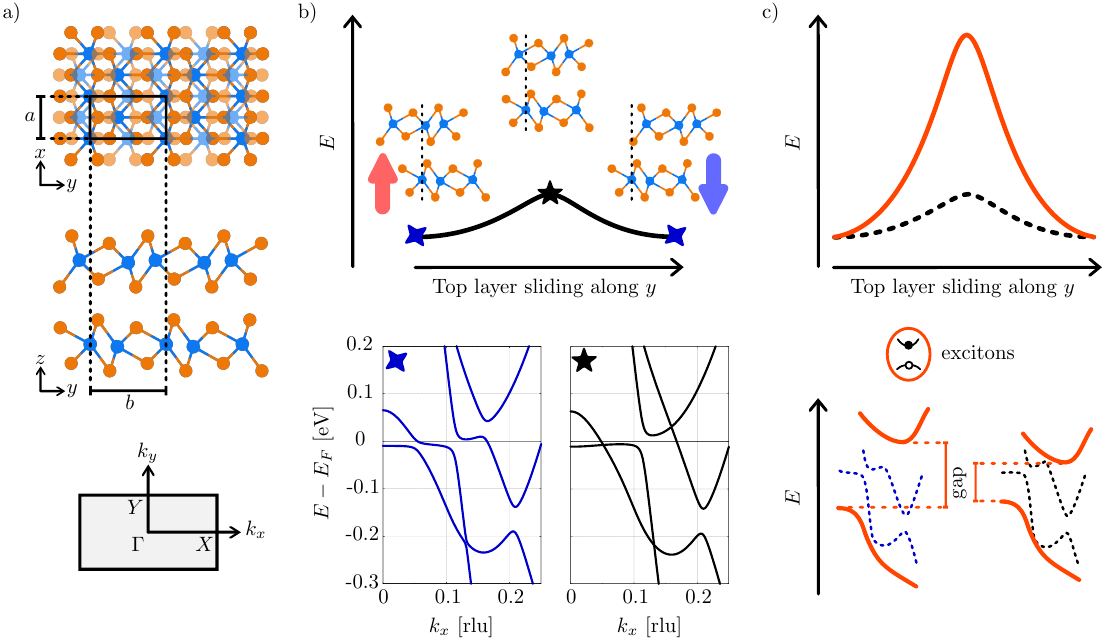}
\caption{
{\bf The potential barrier for the sliding process.} a) Top view, side view and Brillouin zone of bilayer WTe$_2$. In the top view the top layer is shaded for better clarity and the black rectangle marks the unit cell with the corresponding lattice parameters $a$ and $b$. Blue atoms represent W, orange ones Te. b) Top: sketch of the energy barrier of the sliding process. 
The two equivalent ground states have opposite out-of-plane polarization (up/down arrows). The configuration corresponding to the energy maximum is the GMS structure. The top layer displacement from the GMS structure to the ground state structure is largely magnified in this sketch. Bottom: DFT-PBE bands corresponding to the ground state structure (left) and to the GMS structure (right). The bands are plotted along the $\Gamma X$ path and the zero of the energy axis is the Fermi energy, $E_F$. c) Effects of the formation of the excitonic insulator phase (red curves). The sketch shows that tightly bound excitons can modify the band structure (bottom), and possibly open a gap, hence lowering the energy of the system and thus changing the energy barrier for the sliding process (top). The red solid bands represent qualitatively how the DFT bands change after including excitonic effects; the dashed bands, shown as reference, reproduce the DFT-PBE bands of panel b) around $E_F$ for the displaced and GMS structures (left and right, respectively).
}
\label{fig:full_story}
\end{figure*}

The open question has to do with the very mechanism inducing ferroelectricity in this system. There is consensus on the fact that an in-plane sliding along a specific direction ($y$ for bilayer WTe$_2$ as shown in Fig.~\ref{fig:full_story}a) changes the charge distribution between the layers and thus the electric polarization, as initially proposed theoretically \cite{li_binary_2017, wu_sliding_2021}: the sign of the polarization is reversed when the shift takes place in the opposite direction. 
However, while descriptions of this process based on density functional theory (DFT) lead to an interlayer charge transfer compatible with the measured polarization \cite{yang_origin_2018, liu_vertical_2019}, the accurate determination of the sliding energy barrier is still a challenge \cite{tang_sliding_2023, deng_deterministic_2025, gu_2Dferroelectricity_2025}: an issue especially relevant for the room-temperature operation of ferroelectric semimetals.

For ferroelectric semiconductors, like h-BN, the tiny potential barrier ($\sim 2$ meV) is sufficient to stabilize ferroelectricity, since both lattice and electrons are rigid and slide with macroscopic mass \cite{tang_sliding_2023}. For ferroelectric semimetals, the electron system is compressible \cite{Pines1999} and reacts to sliding (Supplementary Section I). As a consequence, the sliding barrier is reduced by at least one order of magnitude ($\sim 0.1$ meV, see Supplementary Figure 1). The suppression of the barrier, which was not appreciated by previous work (DFT estimates of Refs.~\onlinecite{yang_origin_2018, liu_vertical_2019, bai_sub-nanosecond_2025} were significantly larger), calls into question the actual origin of the ferroelectric order, its stability and switching properties.

Here we move from the recent evidence of important excitonic effects in mono- and bi-layer WTe$_2$, leading to a ground-state excitonic insulator (EI) phase \cite{sun_evidence_2022, jia_evidence_2022}, and analyze the role that electron-hole (e-h) interaction can play in the ferroelectric properties of bilayer WTe$_2$. 
This interaction, which may sustain a many-body gap and hence make the electron system incompressible, was not taken into account in previous theoretical studies. Note that the possible role of excitons was already remarked in Ref.~\onlinecite{fei_ferroelectric_2018}, based on the observation of identical electron and hole densities in the ferroelectric phase (as well as in Ref.~\onlinecite{jindal_coupled_2023} for MoTe$_2$, but in the superconducting phase).

More generally, excitonic effects are widespread in 2D semiconductors and semimetals, due to enhancement of e-h correlations with reduced dimensionality \cite{Fogler2014,Wang2018}. 
If the exciton binding energy exceeds the semiconductor gap size (or does not vanish in the semimetal), excitons may undergo Bose-Einstein condensation. 
This results into a new ground state, the EI \cite{jerome_excitonic_1967,Halperin1968}, which qualitatively differs from the pristine phase  since a many-body gap opens \cite{DiSalvo1976,Khveshchenko2001,Bao2012,Velasco2012,Kono2017,Varsano2017,Varsano2020,Ataei2021,Rickhaus2021,Gao2023,Gao2024,jia_evidence_2022,sun_evidence_2022,Shi2022,Traum1978,Cercellier2007,Drut2009,Feldman2009,Gamayun2009,Wakisaka2009,Freitag2012,Rodin2013,Sun2015,Wang2020,Ma2021,Burg2018,Liu2022,Qi2025}, and/or a crystal symmetry is spontaneously broken \cite{Halperin1968,DiSalvo1976,Khveshchenko2001,Bao2012,Velasco2012,Kono2017,Varsano2017,Varsano2020,Ataei2021,Rickhaus2021,Gao2023,Gao2024,Kogar2017,Kaiser2018,Bretscher2021} (though concurrent lattice distortions spoil the EI picture in several candidate bulk materials \cite{Nakano2018,Hedayat2019,Yan2019,Zhou2020,Windgatter2021,pashov_tise2_2025}). 
The most fascinating aspect of the EI is the emergence of macroscopic quantum coherence \cite{Halperin1968,Portengen1996b,Eisenstein2004,Rontani2005a,Littlewood2008,Rontani2013,Cooper2020}, which may appear e.g.~as counterflow superconductivity, or anomalous enhancement of interlayer tunnelling, in bilayer systems with spatially separated electrons and holes \cite{Burg2018,Liu2022,Qi2025,Eisenstein2004,Spielman2000,Nandi2012,Li2017,Liu2017}. 
An intriguing possibility is the `excitonic ferroelectricity', where the macroscopic electric dipole arises from  the coherent superposition of the interband electric dipoles associated with condensed excitons \cite{Portengen1996b,Varsano2020,Ataei2021,zheng_2023} (see also Supplementary Section IIC). 
Furthermore, excitons might trigger the transition to superconductivity upon a tiny amount of doping, like in the phase diagram of monolayer WTe$_2$ \cite{Wu2024}, mediate (unconventional) superconductivity \cite{Little1964,Ginzburg1968,Bardeen1973,Sham1983,Sham1984,Littlewood1990,vanWezel2010,Laussy2010,Ilani2016,Crepel2022} and magnetism \cite{Zhitomirsky1999,Liu2025}, and affect topological properties of 2D materials \cite{Varsano2020,Hossain2025}.

In this Communication, after analyzing the uncorrelated ground state of bilayer WTe$_2$ from first principles, we develop a model to account for e-h interactions forming tightly-bound excitons, in order to calculate how they modify the electronic structure of the system and in turn estimate how this affects its energy.  
We find that the bilayer can undergo a transition to a ground-state excitonic phase: The repulsion of the renormalized conduction and valence bands accounts for the energy cost of exciton ionization and makes the bilayer either semimetal or insulating. 
In all events, the energy of the system is lowered. 
This process induces a relevant change in the energy barrier for the sliding process, thus showing that excitonic effects can contribute significantly to the energetics of the ferroelectric switching.

Our work indicates that electronic correlations and a theoretical description beyond DFT can be relevant for 2D vdW systems, even for phenomena such as sliding ferroelectricity where they are typically neglected. Importantly, it indicates that such contribution can stabilize sliding ferroelectricity beyond previous predictions, which increases the conceptual and technological interest of this phenomenon in the broader class of 2D vdW materials.  

Figure~\ref{fig:full_story} summarizes the main ideas that we will discuss in the following paragraphs. The structure of a WTe$_2$ bilayer is shown in Fig.~\ref{fig:full_story}a, in particular the non polar configuration that we label with a black star in the rest of Fig.~\ref{fig:full_story}. We refer to this configuration as \textit{glide-mirror symmetric} (GMS) structure. The next two columns of the figure respectively schematize the picture emerging from single-particle DFT calculations (Fig.~\ref{fig:full_story}b) and from a description that includes excitonic effects (Fig.~\ref{fig:full_story}c).

%%%%%%%%%%%%%%%%%%%%%%%%%%%%%%%%%%%%%%%%%%%%%%
%%%%%%%%%%%%%%%%% RESULTS %%%%%%%%%%%%%%%%%%%%
\vspace{0.2cm}
\noindent \textbf{Results} \\
\textbf{Starting from the DFT picture}\\
Let us first consider the picture obtained by means of a first-principles description at the DFT level. 
When one of the WTe$_2$ layers 
of the GMS structure, say the top one, is shifted horizontally with respect to the bottom layer, the energy of the system is lowered. 
The new configuration, marked by a blue diamond in Fig.~\ref{fig:full_story}b, is characterized by a finite polarization perpendicular to the layers. 
Because of the symmetry of the crystal structure, in bilayer WTe$_2$ this sliding mechanism connects two energetically equivalent ground states with opposite out-of-plane polarization. 
We estimated the energy barrier for this process, sketched in Fig.~\ref{fig:full_story}b, by means of a DFT nudged elastic band (NEB) calculation, obtaining a value of $\sim$ 0.1 meV. Our careful extrapolation of NEB data to zero temperature (Supplementary Section I) provides an estimate that is a small fraction of previous results \cite{yang_origin_2018,liu_vertical_2019,bai_sub-nanosecond_2025}.

\begin{figure}
\includegraphics[width=1.0\linewidth]{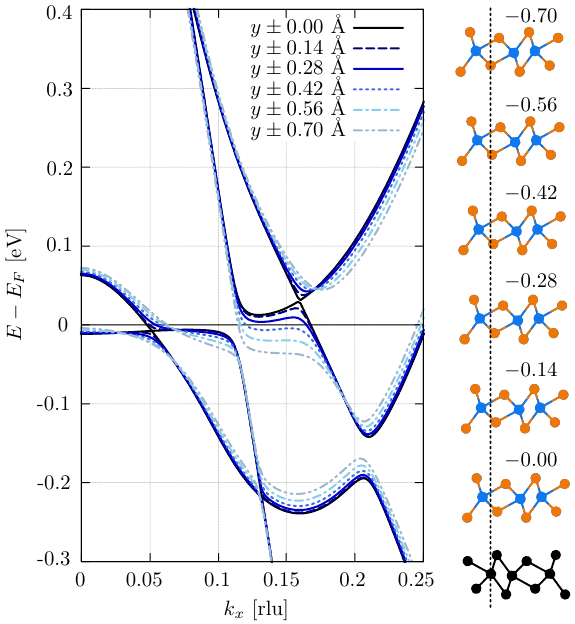}
\caption{
{\bf Effect of layer sliding on band structure.} DFT-PBE bands for different displacements of the top layer along the sliding direction. The plot shows the cut along $\Gamma X$  (thus $k_y=0$); $k_x$ is in units of the corresponding reciprocal lattice vector (rlu). The bottom layer (dyed in black) is fixed and $0.00$ displacement of the top layer corresponds to the GMS structure. Images on the right show the displacements of the top layer (coloured) in real scale, with the dashed line indicating the fixed position of the bottom layer; negative values of the displacements are indicated in {\AA}, as in the key; positive displacements -- not shown -- are analogous and give the same bands.
}
\label{fig:bands_vs_sliding}
\end{figure}

Figure~\ref{fig:full_story}b also shows the band structure for both unpolarized and ground-state polarized configurations, as obtained by DFT calculations within the generalized gradient approximation for the exchange-correlation functional. 
Note that spin-orbit coupling (SOC) is not included in these calculations (see Supplementary Section IV for a discussion), as it leads to minor corrections for the natural stacking of the layers considered here, at difference with geometries that preserve the inversion symmetry \cite{muechler_topological_2016}.   
A systematic display of band structure results for increasing relative displacement of the layers is given in Fig.~\ref{fig:bands_vs_sliding}. 
Among these values, the displacement $\bar{d} = 0.28$ {\AA} gives the lowest total energy in the DFT calculation. The corresponding bands are represented by a solid blue line, as in Fig.~\ref{fig:full_story}b. 
Note that, at this level of description, the system is semimetallic. 
We expect the gap opening effect of DFT hybrid functionals \cite{zheng_quantum_2016, wang_ferroelectric_2019} essentially not to alter the predicted energy barrier.

The comparison of DFT-level calculations with experiments is intriguing. 
On one side, the charge transfer between the layers, as computed from the ground state charge density, gives an areal polarization density of $1.5 \times10^{4}$ e$\,\cdot\,$cm$^{-1}$, in agreement with the experimental measurements \cite{fei_ferroelectric_2018} and previous theoretical results \cite{yang_origin_2018, liu_vertical_2019}. 
On the other side, the metallic character resulting from the calculated band structure is at odds with the measured transport gap of a few meV \cite{sun_evidence_2022}. 
This hints at a polarization density and energy barrier higher than predicted, as DFT overestimates the electronic screening. \\

\vspace{0.2cm}
\noindent\textbf{Beyond DFT: including excitonic interactions} \\
In the following we analyze how this picture is modified by the effects of e-h interaction, to assess the possible contribution of an excitonic phase to ferroelectricity. 
To make the problem tractable, we introduce some approximations. 
We consider spinless excitons, as spin effects are unlikely to affect ferroelectricity (see Supplementary Section IV for further discussion). 
Furthermore, we take that all condensing excitons have zero momentum, $\bm{q}_x=0$ (direct transitions). 
Otherwise, the excitons would sustain a charge density wave, with period related to the momentum $\bm{q}_x$. 
Since for all possible choices of $\bm{q}_x$ the charge modulation would encompass at least two lattice units, the net coupling of the density wave with the macroscopic electronic dipole would likely average out. 

A full first-principles treatment is beyond our scope, as predictive accuracy requires computational tasks that are demanding for a semimetal, such as the calculation of the GW correction to band structure including dynamical screening effects, as well as the evaluation of the intraband contribution to electronic polarizability 
\cite{Liang2015, leon_frequency_2021, champagne_quasiparticle_2023, leon_efficient_2023, guandalini_efficient_2024, sesti_efficient_2025}.
Therefore, we rely on the envelope function approximation, which allows us to both solve the exciton problem and compute the renormalized band structure of the excitonic insulator, building on DFT results. Importantly, we model the e-h attraction through the Rytova-Keldysh potential \cite{Rytova_screened_1967, Keldysh_Coulomb_1979, cudazzo2011dielectric}, which parametrizes the interaction strength through a single parameter, the 2D polarizability of the bilayer, $\alpha_{\mathrm{2D}}$, as defined in Eq.~\eqref{eq:Ryota}.   

We start by writing the (Bethe-Salpeter) equation of motion of excitons based on the
DFT-PBE band structure of Fig.~\ref{fig:bands_vs_sliding}. 
Whereas our framework works for an arbitrary number of bands, here we limit their number to two for the sake of simplicity (results from a four-band model are presented in Supplementary Section II and shown for comparison in Supplementary Figure 4).
We  use the highest valence ($i=1$) and lowest conduction ($i=2$) bands of Fig.~\ref{fig:bands_vs_sliding}, which are separated by gaps at $k_x \approx 0.12$ rlu and $k_x \approx 0.21$ rlu, respectively, as the eigenvalues $\varepsilon_i(\bm{k})$ of the noninteracting Hamiltonian $H_0(\bm{k})$, a $2\times2$ diagonal matrix in $\bm{k}$ space. 
The two bands, which are partially filled in the semimetal ground state predicted by DFT, become either occupied or empty in the excitonic insulator phase. 
The eigenvalue problem for the excitons is 
\begin{equation}
\label{eq:BSE_eigenval_prob}
    \sum_{\bm{k}'} H_{\bm{k}}^{\bm{k}'} \Psi_x^{\bm{k}'}
    =
    E_x \Psi_x^{\bm{k}} \, ,
\end{equation}
with the exciton Hamiltonian matrix element,  
$ H_{\bm{k}}^{\bm{k}'}$, being
\begin{equation}
\label{eq:BSE_mat_el}
    H_{\bm{k}}^{\bm{k}'} = 
    [\varepsilon_{2}(\bm{k}) - \varepsilon_{1}(\bm{k})] \delta_{\bm{k},\bm{k}'} -W(\bm{k}-\bm{k}') \, .
\end{equation}
Here $\Psi_x^{\bm{k}}$ is the exciton wave function, i.e., the probability amplitude for exciting an electron from valence to conduction band conserving its momentum $\bm{k}$, $E_x$ is the exciton energy, and
$W(\bm{q})$ is the e-h attraction that transfers momentum $\bm{q}$ between e-h pairs. 
In view of the postulated exciton condensation, we assume bands 1 and 2 to be respectively filled and empty, hence $\varepsilon_{2}(\bm{k}) - \varepsilon_{1}(\bm{k})$ is always positive. 
The explicit form of $W$ \cite{Rytova_screened_1967, Keldysh_Coulomb_1979, cudazzo2011dielectric} is
\begin{equation}
\label{eq:Ryota}
    W(q) = \frac{e^2}{2\varepsilon_0 A} \frac{1}{q(1+2\pi\,\alpha_{\mathrm{2D}}\,q)} \, ,
\end{equation}
where $A$ is the unit cell area and $\varepsilon_0$ the vacuum permittivity. 

\begin{figure}
\includegraphics[width=1.0\linewidth]{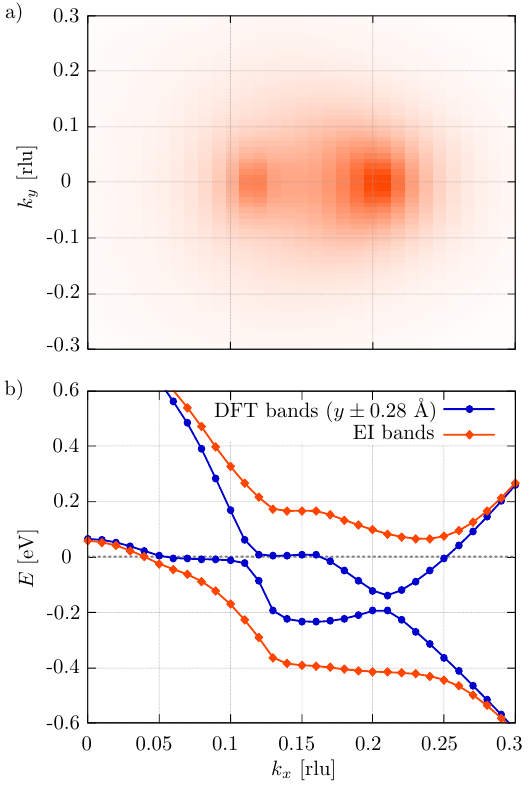}
\caption{
{\bf Excitons and excitonic insulator.} Results for the bilayer structure with sliding displacement $y = \bar{d} = \pm 0.28$ {\AA} and $\alpha_{\mathrm{2D}} = 4.3$ {\AA}. a) Exciton wave function in $\bm{k}$ space. The contour plot represents the probability amplitude $\Psi_x^{\bm{k}}$ of exciting an e-h pair by transferring an electron from the valence to the conduction band state at fixed momentum $(k_x,k_y)$. The range of the color scale is $[0,0.11]$. b) Condensation of excitons leads to the reconstruction of the bands (EI) with the opening of a gap. The plot shows the cut for $k_y = 0$. The EI indirect band gap is $7$ meV.
}
\label{fig:exc_wf_and_ei_bands}
\end{figure}

Figure~\ref{fig:exc_wf_and_ei_bands}a plots the contour map of $\Psi_x^{\bm{k}}$ obtained for the equilibrium layer displacement $\bar{d}$ and e-h attraction strength sufficient to open a gap in the excitonic phase, $\alpha_{\mathrm{2D}} = 4.3$ {\AA} (see Supplementary Section III for further discussion). 
This choice gives a tiny gap of 3 meV in the GMS structure. 
The resulting binding energy is $256$ meV, smaller than the one we  obtained for the monolayer, $\sim$ 300 meV \cite{sun_evidence_2022} (Supplementary Figure 6). 
This is reasonable, as the electronic screening should be more effective in the bi- than in the mono-layer. 
 The maximum amplitude of $\Psi_x^{\bm{k}}$ is reached at the two locations in the Brillouin zone region shown in Fig.~\ref{fig:exc_wf_and_ei_bands}a that correspond to the two energy gaps
along the $\Gamma X$ cut, where the excitation energy of free e-h pairs is lower. 
Note that $\Psi_x^{\bm{k}}$ may be regarded as a proxy of the macroscopic wave function of the exciton condensate \cite{Kohn1967b} that we evaluate below, the discrepancies arising from Pauli blockade effects when condensing excitons fill in the phase space. Therefore, we anticipate that the Brillouin zone region where the band renormalization due to the condensate is stronger approximately mirrors the map of $\Psi_x^{\bm{k}}$ (Fig.~\ref{fig:exc_wf_and_ei_bands}b). 

\vspace{0.2cm}
\noindent\textbf{Contribution of exciton condensation to the ferroelectric transition} \\
We next investigate whether the formation of bound e-h pairs, the excitons, can be energetically favorable in the initial semimetal state, such that the system will  undergo a transition to the  `excitonic' insulator (EI) \cite{Keldysh1964, jerome_excitonic_1967, Halperin1968} or semimetal \cite{Zittartz1967} phase. 
In EIs, excitons form spontaneously, i.e., in the absence of any photoexcitation, condense, and collectively sustain the reconstruction of the bands and the onset of a gap. 
A condition for energy gain in the EI formation is a large exciton binding energy, which in some 2D systems can be achieved thanks to the reduced screening of the e-h Coulomb interaction, as we have demonstrated for WTe$_2$ monolayer \cite{sun_evidence_2022}. 

To explore the possibility of this kind of transition in bilayer WTe$_2$ driven by the formation of direct excitons, we extend the model of Ref.~\onlinecite{jerome_excitonic_1967} to our case, which in principle includes several bands.
The excitonic insulator state is written as
\begin{equation}
    |\Psi_{\mathrm{EI}}\rangle =
    \prod_{i\bm{k}} \hat{\alpha}^{\dagger}_{i\bm{k}} |\mathrm{vac}\rangle \, ,
\end{equation}
where the EI valence band annihilation operators $\hat{\alpha}_{i\bm{k}}$ can be expressed as
\begin{equation}
    \hat{\alpha}_{i\bm{k}} = \sum_j a_{ij\bm{k}} \hat{a}_{j\bm{k}} \, 
\end{equation}
and $|\mathrm{vac}\rangle$ is the vacuum state with no electrons. 
In this last expression $\hat{a}$ is either a conduction or valence band Bloch state of the pristine semimetal and the coefficients $a_{ij}$ form a unitary matrix, to be determined self-consistently for each $\bm{k}$ point. 
The reconstructed bands $E_{i}(\bm{k})$ are then obtained from the diagonalization of
\begin{equation}
\label{eq:mat_2x2_Delta}
    \begin{pmatrix}
        \varepsilon_1(\bm{k}) & 
        \Delta(\bm{k}) \\
        \Delta(\bm{k}) & 
        \varepsilon_2(\bm{k})
    \end{pmatrix} \, ,
\end{equation}
where $\Delta(\bm{k})$ is the (single) gap function for the special case of two bands. Its multiband form is
\begin{equation}
   \Delta_{ij}(\bm{k}) =  
   \sum_{l,l',\bm{k}'} W(\bm{k}-\bm{k}')
  % \langle \psi_i(\bm{k})|\psi_{i'}(\bm{k}')\rangle
  % \langle \psi_j(\bm{k})|\psi_{j'}(\bm{k}')\rangle \\
   \ [a_{\bm{k}'}^{-1}]_{jl} [a_{\bm{k}'}^{-1}]_{il'}
   \langle \hat{\alpha}^{\dagger}_{l\bm{k}'} \hat{\alpha}_{l'\bm{k}'}\rangle_{\Psi_{\mathrm{EI}}} \, .
\end{equation}
Supplementary Section II details the analysis in the case of four bands.  
Since the $\Delta$'s depend on the eigenvalues $E_i(\bm{k})$ of the matrix above through the expectation values 
\begin{equation}
    \langle \hat{\alpha}^{\dagger}_{l\bm{k}'} \hat{\alpha}_{l'\bm{k}'}\rangle_{\Psi_{\mathrm{EI}}} = 
    \delta_{l,l'} f(E_l(\bm{k}'))
\end{equation}
as well as its eigenvectors $a_{\bm{k}}$, the problem has to be solved self-consistently (here the Fermi-Dirac function $f$ is $1$ and $0$ for valence and conduction). 
In the first step of the self-consistent cycle the sum over $l$ and $l'$ is replaced by $\Psi_x^{\bm{k}'}$, the (lowest energy) exciton wavefunction from the solution of Eq.~\eqref{eq:BSE_mat_el}. 
The final, reconstructed bands $E_i(\bm{k})$ come from the diagonalization of the matrices in Eq.~\eqref{eq:mat_2x2_Delta} with the converged values of $\Delta(\bm{k})$. 

We are now ready to show results for the EI phase, as schematically summarized in Fig.~\ref{fig:full_story}c. The calculated EI reconstructed bands with the indirect narrow gap $\approx 7$ meV, comparable to the measured one \cite{sun_evidence_2022},
are given in Fig.~\ref{fig:exc_wf_and_ei_bands}b for the equilibrium relative layer shift $\bar{d}$, compared to the starting DFT bands.

\begin{figure}
\includegraphics[width=1.0\linewidth]{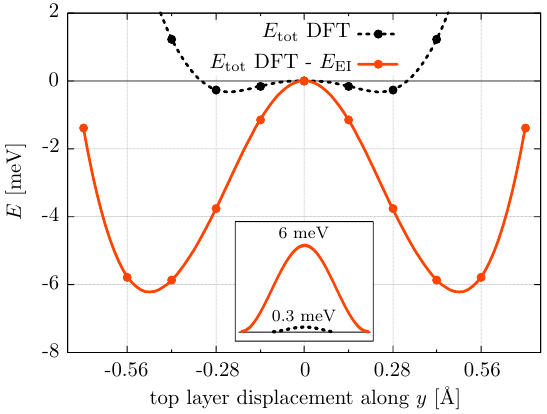}
\caption{
\textbf{Excitonic renormalization of the sliding energy barrier.} Energy versus sliding displacement, with (solid red) and without (dashed black) e-h interaction effects. The dashed black curve is the DFT total energy; the solid red curve is obtained by subtracting the corresponding $E_{\mathrm{EI}}$ value. For both curves the energy zero is set at $d=0$ displacement (GMS configuration). The interpolation lines are a guide to the eye. The inset shows, in scale, the variation of the energy barrier, which increases from $0.3$ to $6$ meV.
}
\label{fig:dft_vs_dftei_barrier}
\end{figure}

To estimate the energy gain in the EI phase for all the top layer displacements, we start from the plot of the DFT total energy $E_{\mathrm{tot}}$ as a function of the sliding along $y$ (Fig.~\ref{fig:dft_vs_dftei_barrier}, black dashed line). 
Without considering the excitonic effects, the energy difference between displacements $0$ and $\bar{d}$ is $\approx 0.3$ meV, which is comparable with the NEB barrier. 
We emphasize that the dashed black curve of Fig.~\ref{fig:dft_vs_dftei_barrier} and the NEB barrier have diverse nature: The latter is obtained through force optimizations determined by the NEB algorithm, whereas the former is the difference in DFT total energy for given layer displacement.

We now evaluate the total energy gain due to exciton condensation, $E_{\mathrm{EI}}$, as
\begin{equation}
\label{eq:E_EI}
    E_{\mathrm{EI}} = 
    \frac{2}{N_{\bm{k}}}
    \sum_{\bm{k}} [\varepsilon_{1,\mathrm{occ}}(\bm{k}) + \varepsilon_{2,\mathrm{occ}}(\bm{k}) - E_{1}(\bm{k})] + \frac{E_{\mathrm{gap}}}{2} \, .
\end{equation}
Here $\varepsilon_{i,\mathrm{occ}}(\bm{k})$ are the occupied (i.e. $\varepsilon \leq E_\mathrm{F}$) DFT energy bands, $E_1(\bm{k})$ is the reconstructed valence band energy for the EI phase, lowered by $-E_\mathrm{gap}/2$ (at zero temperature the chemical potential is in the middle of the gap), $N_{\bm{k}}$ is the total number of k-points sampling the Brillouin zone, and the factor $2$ accounts for spin degeneracy. 
Note that Eq.~\eqref{eq:E_EI} holds only for the insulating phase. 
For the semimetallic excitonic phase, 
$E_{\mathrm{gap}}=0$, and $E_{1}(\bm{k})$ in Eq.~\eqref{eq:E_EI} must be replaced with
$E_{1,\mathrm{occ}}(\bm{k})+E_{2,\mathrm{occ}}(\bm{k})$, where $E_\mathrm{F}$ has to be determined self-consistently, for each step of the iterative cycle. It is then immediate to check that $E_{\mathrm{EI}} = 0$ in the absence of excitonic effects. 

Importantly, the variations in the starting DFT bands and their anticrossings (Fig.~\ref{fig:bands_vs_sliding})  impact the final $E_i(\bm{k})$ bands and $E_{\mathrm{gap}}$ significantly. 
It follows that the energy gain $E_{\mathrm{EI}}$ is highly sensitive to the sliding displacement value $d$ along $y$, as clear from Fig.~\ref{fig:dft_vs_dftei_barrier} (solid red curve). 
Namely, the energy difference between the ferroelectric structure ($d=\bar{d}$ displacement) and the GMS structure ($d=0$) with no dipole is modified from $0.3$ meV to $\approx4$ meV. 
For a displacement of $0.4$ {\AA} this value increases up to around $6$ meV . 
The value of the areal polarization density at the minima of the renormalized barrier is $2\times10^4$ e$\,\cdot\,$cm$^{-1}$, which compares to the experimental estimate of $1\times10^4$ e$\,\cdot\,$cm$^{-1}$ \cite{fei_ferroelectric_2018}. 
Overall, the energy barrier for the sliding process is heavily renormalized by condensation, due to the sensitivity of exciton binding to the modifications of the band structure.

%%%%%%%%%%%%%%%%%%%%%%%%%%%%%%%%%%%%%%%%%%%%%%
%%%%%%%%%%%%%%%% DISCUSSION %%%%%%%%%%%%%%%%%%
\vspace{0.2cm}
\noindent\textbf{Discussion} \\
Our theory shows that a gap opening in bilayer WTe$_2$ is induced by the condensation of direct excitons. 
Whereas the absolute value of the indirect gap depends on the screening parameter $\alpha_{\mathrm{2D}}$, the gap size variation (and in turn $E_{\mathrm{EI}}$) with the layer shift coordinate $d$ does not, as illustrated from Supplementary Figure 7 and discussed in Supplementary Section III. 
Therefore, our estimate of the `excitonic' barrier height should be robust against the uncertainty associated with the simplified approach. 

In conclusion, we have pointed out that e-h interaction effects can play an important role in sliding ferroelectricity that is not captured by a first-principles description at the DFT level. 
For the WTe$_2$ prototype bilayer we show that, as long as the exciton coherence survives, relevant modifications result in the energetics of the system that contribute to stabilize ferroelectricity upon sliding. 
This will add to other effects, both intrinsic (intralayer stiffness, vdW interaction, quantum and thermal fluctuations) and extrinsic (e.g.~domain wall dynamics, defects, boundaries, etc), dictating the Curie temperature and switching properties of the system.

We expect that relevant excitonic contributions will be present in general throughout 2D systems, where e-h Coulomb interactions are greatly enhanced by the reduced screening. 
Combined with the easy relative sliding typical of van der Waals layers, these results indicate that 2D sliding ferroelectricity may be even more widespread and robust than previously expected.

%%%%%%%%%%%%%%%%%%%%%%%%%%%%%%%%%%%%%%%%%%%%%%
%%%%%%%%%%%%%%%%% METHODS %%%%%%%%%%%%%%%%%%%%
\vspace{0.2 cm}
\noindent{\textbf{Methods}} \\
DFT and NEB calculations are performed using the Quantum ESPRESSO package \cite{giannozzi_advanced_2017}. 
We adopt the Perdew-Burke-Ernzerhof (PBE) approximation for the exchange-correlation functional \cite{perdew_generalized_1996} and norm-conserving pseudopotentials \cite{hamann_optimized_2013}. 
Unless otherwise specified, the cutoff value for plane wave expansion of wave functions is $100$ Ry and the Brillouin zone is sampled by a $22\times12$ Monkhorst-Pack grid of k-points \cite{monkhorst_special_1976}. 
Van der Waals interactions are taken into account through the Grimme-D3 method \cite{grimme_a_consistent_2010} with Becke-Johnson damping \cite{grimme_effect_2011}. 
Spurious interactions between periodic replicas in the out-of-plane direction are avoided by setting the cell dimension along $z$ to $40$ {\AA}. 
NEB calculations employ the climbing image scheme to determine the transition path \cite{henkelman_climbing_2000}, with a $0.01$ eV$\,\cdot\,${\AA}$^{-1}$ threshold on the forces orthogonal to the path. 
The plane wave expansion cutoff in NEB calculations is set to $120$ Ry.
The optimized lattice parameters and atomic positions are obtained by means of a structure relaxation with a threshold of $1$ meV$\,\cdot\,${\AA}$^{-1}$ on atomic forces and a cut-off value of $200$ Ry for the plane wave expansion. 
The resulting values are $3.473$ {\AA} and $6.276$ {\AA} for the $x$ and $y$ cell parameters respectively. 
The model BSE and EI gap equation are implemented in a publicly available custom code \cite{dalessio_2025_17311469}. 
The k-points in the calculations discussed here sample the region of the BZ corresponding to $k_x = [0,0.3]$ and $k_y = [-0.3,0.3]$ in units of the corresponding reciprocal lattice vectors (rlu) with a $30\times60$ mesh. 
In the self-consistent gap equation we check convergence on the values of $\Delta(\bm{k})$: the cycle stops at the $n$-th step if $|\Delta^{[n-1]}(\bm{k}) - \Delta^{[n]}(\bm{k})| < \varepsilon_{\mathrm{thr.}}$, and we set  $\varepsilon_{\mathrm{thr.}} = 0.5$ meV.

%%%%%%%%%%%%%%%%%%%%%%%%%%%%%%%%%%%%%%%%%%%%%%
%%%%%%%%%%%%% ACKNOWLEDGEMENTS %%%%%%%%%%%%%%%
%\begin{acknowledgments}
%\end{acknowledgments}
\vspace{0.2 cm}
\noindent{\textbf{Acknowledgments}} \\
We are grateful to Giacomo Sesti and Claudia Cardoso for many useful discussions. 
Access to high performance computing resources was provided by EuroHPC Joint Undertaking (JU) through the project EHPC-EXT-2022E01-022 and ISCRA, that granted access to the LEONARDO supercomputer, owned by the EuroHPC JU, hosted by CINECA (Italy). 
This work was supported in part by: 
the MaX -- MAterials design at the eXascale -- European Centre of Excellence, co-funded by the EuroHPC JU and Ministero delle Imprese e del Made in Italy (grant agreement No. 101093374); 
ICSC -- Centro Nazionale di Ricerca in High Performance Computing, Big Data and Quantum Computing -- funded by the European Union through the Italian Ministry of University and Research under PNRR M4C2I1.4 (grant CN00000013).

%%%%%%%%%%%%%%%%%%%%%%%%%%%%%%%%%%%%%%%%%%%%%%
%%%%%%%%%%%%%%%% SUPP. INFO %%%%%%%%%%%%%%%%%%
\clearpage

\onecolumngrid
\renewcommand{\figurename}{Supplementary Figure}
\setcounter{figure}{0}

\begin{center}
{\large \textbf{Supplementary Information}}
\end{center}

%%%%%%%%%%%%%%%%%%%%%%%%%%%%%%%%%%%%%%%%%%%%%%
%%%%%%%%%%%%%%%%% SMEARING %%%%%%%%%%%%%%%%%%%
%\clearpage
\section{Estimate of the DFT-NEB ferroelectric switching barrier}

In this section we apply the nudged elastic band (NEB) method, as implemented in the DFT framework \cite{giannozzi_advanced_2017}, to estimate the energy barrier of the sliding ferroelectric switching.
We take special care of exploiting the Fermi-Dirac occupation function to account for the electronic temperature through the smearing parameter, and discuss how this impacts the  NEB energy barrier, as well as the DFT band structure. 

In the following we report results for a smearing parameter decreasing from $5\times10^{-3}$ Ry (i.e., $\approx 800$ K) down to $5\times10^{-4}$ Ry ($\approx 80$ K).
Though $800$ K may look like a large value, this is the typical order of magnitude used to speed up the convergence of the self-consistent DFT calculation, particularly when dealing with a coarse $\bm{k}$-point sampling of the Brillouin zone combined with a Fermi-Dirac occupation function. 
As a lower smearing parameter requires 
a denser Brillouin-zone sampling to achieve numerical convergence, we increased the $\bm{k}$-point grid from $22\times12$ to $66\times36$, which yields fully converged results.

Furthermore, as the interplay between $\bm{k}$-point density and the smearing parameter influences the equilibrium configuration through the evaluation of forces, we  performed variable-cell relaxations to optimize both the atomic positions and the lattice parameters 
for each given combination of smearing parameter and $\bm{k}$-point grid.

\begin{figure}[h]
\includegraphics[width=0.85\linewidth]{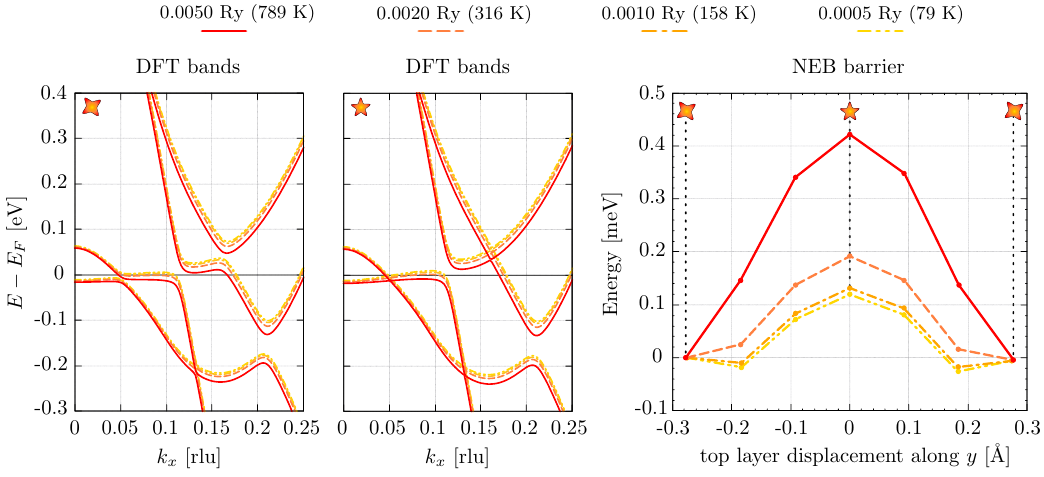}
\caption{
{\bf DFT and NEB results at different electronic temperatures.} 
DFT band structures of bilayer WTe$_2$ (left and center panels) and NEB energy barrier of the sliding ferroelectric switching (right panel) using a $66\times 36$ {\bf k}-point grid for the different smearing parameters reported on top. The smearing parameter fixes the electronic temperature through the Fermi-Dirac occupation function. 
The band structures in the left panel correspond to the equilibrium configuration obtained via variable-cell relaxation (diamond symbol), whereas those in the center panel to the glide-mirror symmetric (GMS) structure found by the NEB calculation (star symbol). 
In the left and center panels the bands are shifted by their respective Fermi energy, whereas in the right panel each NEB curve is shifted so that its first image (i.e.~the starting configuration of the reaction path) is at zero energy. The values on the horizontal axis of the right panel correspond to the relative sliding displacement between the layers, with zero displacement corresponding to the GMS configuration. The circles on the NEB curves mark the images used for the discretization of the reaction path, and the connecting lines are a guide to the eye.
}
\label{fig:siaddedfig05}
\end{figure}

As shown in the left and center panels of Supplementary Figure \ref{fig:siaddedfig05}, the smearing parameter changes the DFT band structure---already converged for a $22\times12$ grid---without affecting the magnitude of the gaps at $k_x \approx 0.12$ rlu and $k_x \approx 0.21$ rlu.
Therefore, our results for the excitonic renormalization of the switching barrier, presented in the main text, are unaffected.  
However, the smearing parameter strongly affects the occupation of valence states for $0.05 < k_x < 0.12$ rlu, which are filled only for temperatures higher than $\approx 300$ K. This band region is remarkably flat not only along $k_x$ but also along $k_y$ (cf.~Supplementary Figure \ref{fig:siaddedfig01}c), hence providing a sizable contribution to the density of states in the Fermi energy window. Similarly, a significant enhancement of the density of states originates from the conduction band for $0.12 < k_x < 0.17$ rlu, though at higher temperature and sliding displacement (cf.~Figure 2 of the main text). 

The above behavior points to a wide variation of Gibbs free energy (the extensive counterpart of the chemical potential) with temperature, and reflects in the extreme sensitivity of the NEB potential barrier to the smearing parameter. 
This is illustrated in the right panel of Supplementary Figure \ref{fig:siaddedfig05}, where the energy zero for each curve is fixed to the energy of the respective starting configuration.
For given layer displacement coordinate,
the NEB energy potential increases with the smearing parameter, the maximum variation
corresponding to zero displacement (GMS structure). This may be easily rationalized, as the valence band of the GMS configuration is almost perfectly flat for a significant $k_x$ range (center panel of Supplementary Figure \ref{fig:siaddedfig05}), and hence its filling is most sensitive to temperature. 

This results in a general reduction of the switching barrier with decreasing temperature. 
The trend shown in the right panel of Supplementary Figure \ref{fig:siaddedfig05} suggests that, at zero temperature, the barrier stabilizes to a value around $0.1$ meV. 
The barrier ``width''---the horizontal sliding displacement that connects the two ferroelectric ground states with opposite dipole---shows a similar trend, i.e.,  it shrinks as the electronic temperature is lowered.

The barrier height values previously published were obtained with the same methodology but were significantly higher: $0.6$ meV \cite{yang_origin_2018,bai_sub-nanosecond_2025} and $0.3$ meV \cite{liu_vertical_2019}. 
We note that in those works smaller $\bm{k}$-point grids were used. In conclusion, the huge suppression of the NEB
potential barrier height we report may be ascribed to the very high value of the electronic density of states at Fermi energy, i.e., the extreme compressibility of Fermi liquid. We expect no such effect for the incompressible Fermi system of a large-gap semiconductor, like h-BN. 

Interestingly, the increase in the switching barrier with temperature was recently measured (above 100 K) in bilayer WTe$_2$ and attributed to the appearance of the Fermi surface \cite{bai_sub-nanosecond_2025}. The experiment might be related to the results reported in the main text, as the predicted excitonic indirect gap varies between $3$ meV ($\approx 35$ K, GMS structure) and $7$ meV ($\approx 80$ K, $\bar{d}$ displacement) with sliding.

\begin{figure}[h]
\includegraphics[width=0.5\linewidth]{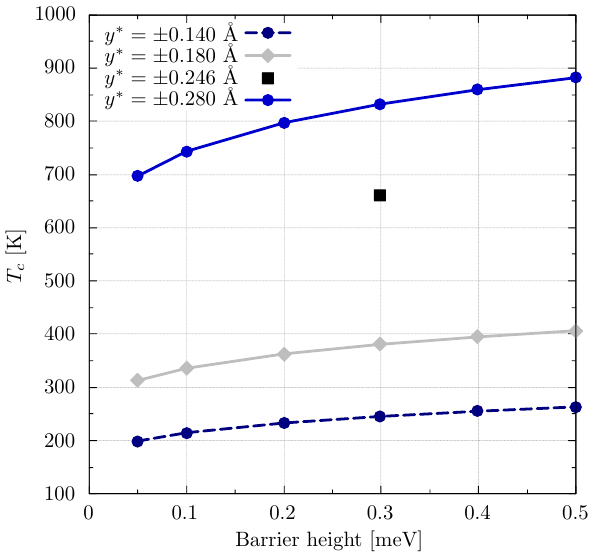}
\caption{
{\bf Ferroelectric switching Curie temperature ($T_c$) after Tang and Bauer (2023).}
The plot reports the numerical solution of equation (19) of Ref.~\onlinecite{tang_sliding_2023}. Here $y^*$ is used to indicate the relative sliding displacement between the GMS configuration and the minimum energy configuration (i.e., half the NEB barrier width). The black square points to the estimate for bilayer WTe$_2$ given in Ref.~\onlinecite{tang_sliding_2023}. The other curves are estimates of $T_c$ for both larger and lower values of barrier height and width, for a range of values relevant to the NEB barrier results of Supplementary Figure \ref{fig:siaddedfig05}. The connecting lines are a guide to the eye.
}
\label{fig:siaddedfig06}
\end{figure}

According to the thermodynamic mean-field theory of Ref.~\onlinecite{tang_sliding_2023}, the key to the high Curie temperature ($T_c$) of sliding ferroelectrics is the rigidity of the layer sliding, which proceeds with a macroscopic mass.
In addition to structural parameters (unit cell area and the 2D Lamé coefficients), 
presumably unchanged in our treatment with respect to the values reported in Table I of Ref.~\onlinecite{tang_sliding_2023}, $T_c$ is most sensitive to the switching
barrier height and width, as shown in Supplementary Figure \ref{fig:siaddedfig06}.
From Supplementary Figure \ref{fig:siaddedfig05} we estimate a barrier height of $\approx 0.1$ meV and width $y^*\approx 0.180$ \AA, leading to $T_c\approx 330$ K, which is half the original prediction for bilayer WTe$_2$ (black square dot \cite{tang_sliding_2023}). As the mean-field treatment of the
layer sliding dynamics is known to overestimate $T_c$ by neglecting quantum fluctuations, it remains unclear whether the DFT treatment alone is able to explain the room-temperature   ferroelectricity of bilayer WTe$_2$. This uncertainty points to the relevance of the excitonic enhancement of the switching barrier height discussed in the main text.

%%%%%%%%%%%%%%%%%%%%%%%%%%%%%%%%%%%%%%%%%%%%%%
%%%%%%%%%%%%%%% 4-BAND MODEL %%%%%%%%%%%%%%%%%
\clearpage
\section{Four-band model}

We develop the four-band model mentioned in the main text to investigate the onset of macroscopic excitonic coherence in bilayer WTe$_2$, as a function of the relative sliding of the two layers, $y$. 
This model fully complies with the space group symmetry of the bilayer for all $y$, in the absence of e-h interaction,
and provides a transparent picture of the orbital character of the exciton wave function. This allows us to analyze the
breaking of microscopic symmetries due to exciton condensation.

\subsection{Noninteracting Hamiltonian}

Within the envelope function approximation, the model provides the two lowest conduction ($c$) and two highest valence ($v$) bands by spanning the Hilbert space through a four-dimensional orbital basis. In this representation, the $\bm{k}$-resolved noninteracting Hamiltonian, $\hat{H}(\bm{k})$, is a $4\times 4$ matrix. The basis is made of the Bloch states at $\Gamma$ obtained \textit{ab initio} for the {\it glide-mirror symmetric} (GMS) configuration ($y=0$), after a rotation. The non rotated states, plotted in Supplementary Fig.~\ref{fig:siaddedfig01}b, are either even or odd with respect to the symmetry operations of the space group, i.e.: (i) the glide-mirror  operation, $\mathbb{G}$, a combination of mirror reflection $z \rightarrow -z$ and half-period translation $ y \rightarrow y+b/2$ along the $y$ axis ($b$ is the lattice constant  along $y$); (ii) the mirror reflection $x \rightarrow -x$, $\mathbb{M}$, with the vertical mirror plane $yz$ intersecting the row of W atoms at the origin.

As apparent from Supplementary Fig.~\ref{fig:siaddedfig01}b, the Bloch states symmetrically spread between the two layers. Since we are interested in the charge transfer and consequent asymmetry between layers, we rotate the basis frame to introduce a layer index, $\sigma_z$, as the eigenvalue of the $2\times 2$ $z$-component Pauli matrix, $\hat{\sigma}_z$,  with eigenvectors $|\chi_{\sigma_z}^{\pm}\rangle$  obeying
$\hat{\sigma}_z|\chi_{\sigma_z}^{\pm}\rangle=\pm |\chi_{\sigma_z}^{\pm}\rangle$.
Here the pseudospinors $|\chi_{\sigma_z}^{+}\rangle$ and $|\chi_{\sigma_z}^{-}\rangle$ are linear combinations of (either conduction or valence) Bloch states, and are localized respectively in the top and bottom layer. 
Furthermore, as the $\mathbb{M}$ symmetry survives for $y\neq 0$, we introduce a second
orbital pseudospin, $\tau_z$, corresponding to orbitals respectively even ($\tau_z=+1$, top row of Supplementary Fig.~\ref{fig:siaddedfig01}b) or odd
($\tau_z=-1$, bottom row) under reflection, i.e., $\hat{\tau}_z|\chi_{\tau_z}^{\pm}\rangle=\pm |\chi_{\tau_z}^{\pm}\rangle$. The complete basis set stems from the external product of the two pseudospinors,  
$|\chi_{\sigma_z}^{\pm},\chi_{\tau_z}^{\pm}\rangle
\equiv |\chi_{\sigma_z}^{\pm}\rangle \otimes  |\chi_{\tau_z}^{\pm}\rangle$.
In this representation, the symmetry operations are
\begin{equation}
    \mathbb{G} = \hat{\sigma}_x \otimes \mathbb{I}_\tau \, ,
    \quad
    \mathbb{M} = \mathbb{I}_\sigma \otimes \hat{\tau}_z \, ,
\end{equation}
where 
$\mathbb{I}_\sigma$ and $\mathbb{I}_\tau$ are the $2\times 2$ unit matrices in the layer and orbital pseudospin spaces, respectively.

The eigenvalues of the noninteracting Hamiltonian for $y=0$,
$\hat{H}_0(\bm{k})$, are---by construction---the DFT energy bands of the GMS structure, 
$\varepsilon_{M,G}^0(\bm{k})$, with 
the subscripts $M=\pm$ and $G=\pm$ labeling the parity of the corresponding eigenvector with respect to $\mathbb{M}$ and $\mathbb{G}$ operations, respectively. 
The explicit form of $\hat{H}_0(\bm{k})$ is   
\begin{equation}\label{eq:H0}
\begin{split}  
    \hat{H}_0(\bm{k}) = &
    \bigg( \frac{\mathbb{I}_\sigma + \hat{\sigma}_x}{2} \bigg)
    \otimes
    \bigg( \frac{\mathbb{I}_\tau + 
\hat{\tau}_z}{2} \bigg)
    \varepsilon_{+,+}^0(\bm{k}) 
    + \bigg( \frac{\mathbb{I}_\sigma - \hat{\sigma}_x}{2} \bigg)
    \otimes
    \bigg( \frac{\mathbb{I}_\tau + \hat{\tau}_z}{2} \bigg)
    \varepsilon_{+,-}^0(\bm{k}) ~+ \\
    & \bigg( \frac{\mathbb{I}_\sigma + \hat{\sigma}_x}{2} \bigg)
    \otimes
    \bigg( \frac{\mathbb{I}_\tau - \hat{\tau}_z}{2} \bigg)
    \varepsilon_{-,+}^0(\bm{k}) 
    + \bigg( \frac{\mathbb{I}_\sigma - \hat{\sigma}_x}{2} \bigg)
    \otimes
    \bigg( \frac{\mathbb{I}_\tau - \hat{\tau}_z}{2} \bigg)
    \varepsilon_{-,-}^0(\bm{k}) \, .
\end{split}
\end{equation}
The energy bands are plotted in Supplementary Fig.~\ref{fig:siaddedfig01}a and c respectively along the $\Gamma$X cut and in the rectangular sector of
the Brillouin zone $k_x \in [0,0.35]$, $k_y \in [-0.1,0.1]$. 

For finite displacement $y$, we add to the total Hamiltonian,
$\hat{H}(\bm{k}) = \hat{H}_0(\bm{k}) + \hat{H}_1 $,
the term $\hat{H}_1$, which is the interaction between the electron and the ``frozen'' phonon corresponding to the rigid sliding of the two layers,
\begin{equation}\label{eq:H1}
    \hat{H}_1 = (\hat{\sigma}_z \otimes \mathbb{I}_\tau)\, \delta_{H_1} \, .
\end{equation}
Here the ``phonon'' parameter $\delta_{H_1}$ depends on $y$ but not on $\bm{k}$ (see Supplementary Section IIE for its estimate). 
The term \eqref{eq:H1} breaks the glide-mirror symmetry, as it does not commute with $\mathbb{G}$, and explicitly accounts for the layer polarization
through the average value of $\left< \hat{\sigma}_z \right>$ over the ground state.

\begin{figure}[h]
\includegraphics[width=1.0\linewidth]{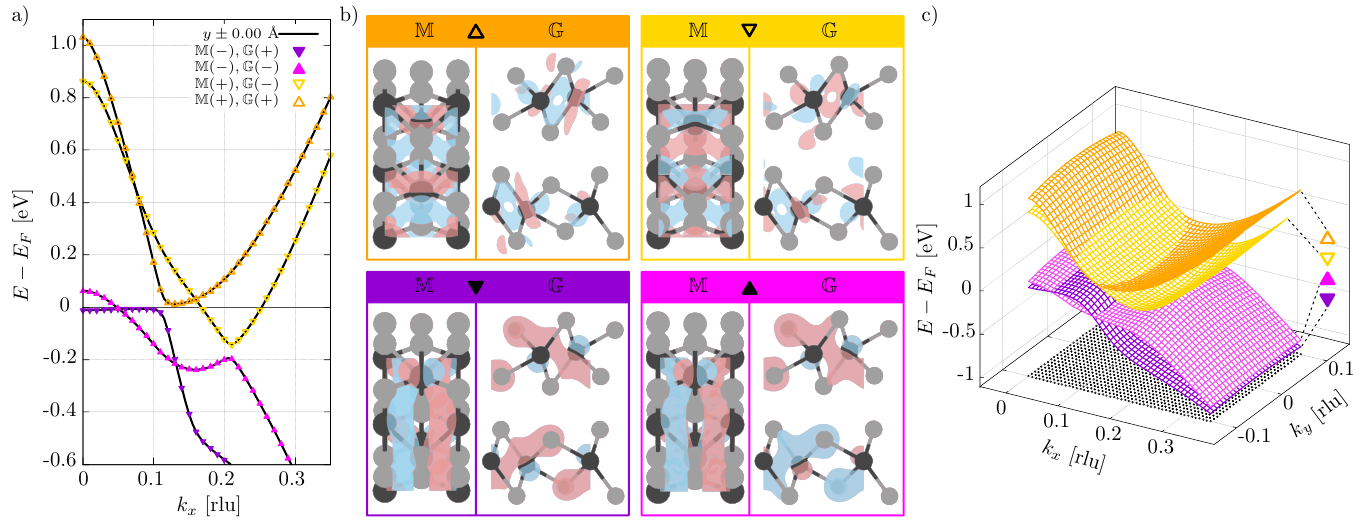}
\caption{
{\bf Orbital basis and symmetry.}
a) Energy bands for the GMS structure along the $\Gamma$X cut of the Brillouin zone computed from density functional theory, as in Fig.~2 of the main text. The superimposed triangle symbols (open/filled and up/down) point to the symmetry classification of the states at $\Gamma$, tabulated in both legend and panel b. Since the glide-mirror symmetry is preserved along the $\Gamma$X line and there are no anticrossings, the symmetry labeling is extended to all $\bm{k}$ points, as the orbital character of each labeled band changes smoothly as one moves through the Brillouin zone. 
b) Wave functions (real part) at $\Gamma$ of the highest two valence and lowest two conduction bands. The amplitude map colour points to the parity of the wave function with respect to mirror ($\mathbb{M}$) and glide-mirror ($\mathbb{G}$) symmetry operations, the red [blue]  corresponding to the positive [negative] amplitude. In each one of the four panels, the left inset shows the top view of the bilayer ($x$ axis in the horizontal direction) and the right inset shows the lateral view ($y$ axis in the horizontal direction). Dark and light gray atoms are W and Te, respectively.
c) Three-dimensional plot of the bands in a selected rectangular range of the Brillouin zone, illustrating the symmetry classification in the $\bm{k}$ space.
}
\label{fig:siaddedfig01}
\end{figure}

\subsection{Bethe-Salpeter equation and Coulomb interaction}

Within the four-band model, the Bethe-Salpeter equation of motion for excitons that generalizes the expression (2) of the main text is
\begin{equation}
\label{eq:BSE_SI}
    \sum_{v',c',\bm{k}'} H_{vc\bm{k}}^{v'c'\bm{k}'}   \Psi_{x}^{v'c'\bm{k}'}
    =
    E_x \Psi_{x}^{vc\bm{k}} \, .
\end{equation}
Now $\Psi_{x}^{vc\bm{k}}$ is the component of the exciton wave function that gives the probability amplitude for exciting an electron from the valence band $v$ to the conduction band $c$, conserving the momentum $\bm{k}$. The indices $v$ and $c$ label two of the four energy bands derived from the solution of $\hat{H}_0+\hat{H}_1$, as defined in \eqref{eq:H0} and 
\eqref{eq:H1}. The kernel of \eqref{eq:BSE_SI} is
\begin{equation}\label{eq:kernelSI}
H_{vc\bm{k}}^{v'c'\bm{k}'} = 
\big[ \varepsilon_c(\bm{k}) - \varepsilon_v(\bm{k}) \big] \delta_{vv'}\,\delta_{cc'}\,\delta_{\bm{k}\bm{k}'}\,
 - \,W_{vc\bm{k}}^{v'c'\bm{k}'} \, ,
\end{equation}
and contains the screened electron-hole Coulomb
attraction, $W$. Importantly, since the model provides us with the layer index $\sigma_z$
for the eigenvectors of $\hat{H}_0(\bm{k})+\hat{H}_1(\bm{k})$,
we can generalize the form of the two-dimensional Rytova-Keldysh potential
to the bilayer system \cite{van_der_donck_interlayer_2018}. Therefore, we resolve the Coulomb interaction matrix element, $W_{vc\bm{k}}^{v'c'\bm{k}'}$,
as
\begin{equation}
\label{eq:Wcvk_SI}
\begin{split} 
W_{vc\bm{k}}^{v'c'\bm{k}'} = &~
 W(\bm{k}-\bm{k}',0) \Big[
 \langle \psi_v^{\mathrm{T}}(\bm{k})|\psi_{v'}^{\mathrm{T}}(\bm{k}')\rangle
 \langle \psi_c^{\mathrm{T}}(\bm{k})|\psi_{c'}^{\mathrm{T}}(\bm{k}')\rangle^*
 + 
 %W(\bm{k}-\bm{k}',0) 
 \langle \psi_v^{\mathrm{B}}(\bm{k})|\psi_{v'}^{\mathrm{B}}(\bm{k}')\rangle
 \langle \psi_c^{\mathrm{B}}(\bm{k})|\psi_{c'}^{\mathrm{B}}(\bm{k}')\rangle^* \Big]
 + \\ &~
 W(\bm{k}-\bm{k}',h) \Big[
 \langle \psi_v^{\mathrm{T}}(\bm{k})|\psi_{v'}^{\mathrm{T}}(\bm{k}')\rangle
 \langle \psi_c^{\mathrm{B}}(\bm{k})|\psi_{c'}^{\mathrm{B}}(\bm{k}')\rangle^* 
 +
 %W(\bm{k}-\bm{k}',h) 
 \langle \psi_v^{\mathrm{B}}(\bm{k})|\psi_{v'}^{\mathrm{B}}(\bm{k}')\rangle
 \langle \psi_c^{\mathrm{T}}(\bm{k})|\psi_{c'}^{\mathrm{T}}(\bm{k}')\rangle^* \Big]
 \, .
\end{split}
\end{equation}
Here $|\psi_i^\mathrm{T}(\bm{k})\rangle$ and $|\psi_i^\mathrm{B}(\bm{k})\rangle$ are the projections of the $i$th eigenvector ($i=c,v$) onto the top and bottom layer, respectively, and the screened interaction is 
\begin{equation} 
    W(\bm{q},z) = \frac{e^2}{2\varepsilon_0 A} \frac{1}{q\,\epsilon(q,z)} \, ,
\end{equation}
with  
\begin{equation}
    \epsilon(q,z) = 
    (1 + 2\pi\,\alpha_{\mathrm{2D}}\,q)\cosh(zq) + 
    [2+
     4 \pi\,\alpha_{\mathrm{2D}}\,q\, 
     (2\pi\alpha_{\mathrm{2D}}q)^2
    ]\sinh(zq)/2 \,. 
\end{equation}
In the above equations, $h=6.9$ {\AA} is the distance between the two layers, chosen as the distance between the planes parallel to the WTe$_2$ sheets passing through the W atoms, 
and $z$ is the vertical coordinate. 
We set $z=0$ ($z=h$) for intralayer (interlayer) interaction.

\subsection{Excitons and spontaneously broken symmetry of the EI phase}\label{s:broken}

The lowest exciton state for the GMS structure is plotted in the left column of Supplementary Fig.~\ref{fig:sifig01}, the four contour maps in $\bm{k}$ plane corresponding to different $vc$ pairs of the wave function component $\Psi_{x}^{vc\bm{k}}$. 
As clear from the colour map, the amplitude with the largest weight is $\Psi_{x}^{23\bm{k}}$ (third panel from top), which qualitatively overlaps with the modulus of the exciton wave function derived within the two-band model. 
The latter is shown in Fig.~3a of main text for finite $y$; the corresponding plot for $y=0$ is qualitatively similar.
The corrections introduced by the four-band model mainly consist in expanding the size of the Hilbert space that spans the exciton state, by means of adding wave function components. 
Note, though, that the weight of new components $\Psi_{x}^{vc\bm{k}}$ decreases with the increase of the e-h excitation energy for $vc$ pair of bands located farther from Fermi energy. 
For the same reason, the maximum weight of $\Psi_{x}^{23\bm{k}}$ occurs in $\bm{k}$ space close to the position of the gaps separating bands 2 and 3.

Since the total momentum of the exciton is zero, the state must transform under symmetry operations according to the space group at $\Gamma$. 
It may be easily recognized from Supplementary Fig.~\ref{fig:sifig01} that the exciton wave function is even under the glide-mirror operation and odd under specular reflection (see for example the parity of e-h pairs at $\Gamma$). 
We checked that, for all $y\neq 0$, the lowest exciton state remains odd under reflection, whereas the glide-mirror symmetry is lifted by the sliding. This behavior rules the broken symmetry of the EI phase, since
the noninteracting ground state wave function, which is even under specular reflection, is distorted by the hybridization with the (odd) exciton wave function at the onset of condensation \cite{Kohn1967b}. 
It follows that the excitonic insulator must spontaneously break the mirror symmetry $x \rightarrow -x$, which is otherwise preserved in the absence of e-h attraction, for all $y$. This hints at ``excitonic ferrolectricity'' \cite{Portengen1996b}, which is driven by the macroscopic condensation of the interband exciton dipoles, here aligned along the $x$ axis. 
Note that such exciton macroscopic dipole is orthogonal to the vertical macroscopic polarization that is observed experimentally. 
We expect the inclusion of spin-orbit coupling in the theory not to alter this picture, but possibly lead to additional symmetry breakings of magnetic type.          
 
\begin{figure}[h]
\includegraphics[width=0.6\linewidth]{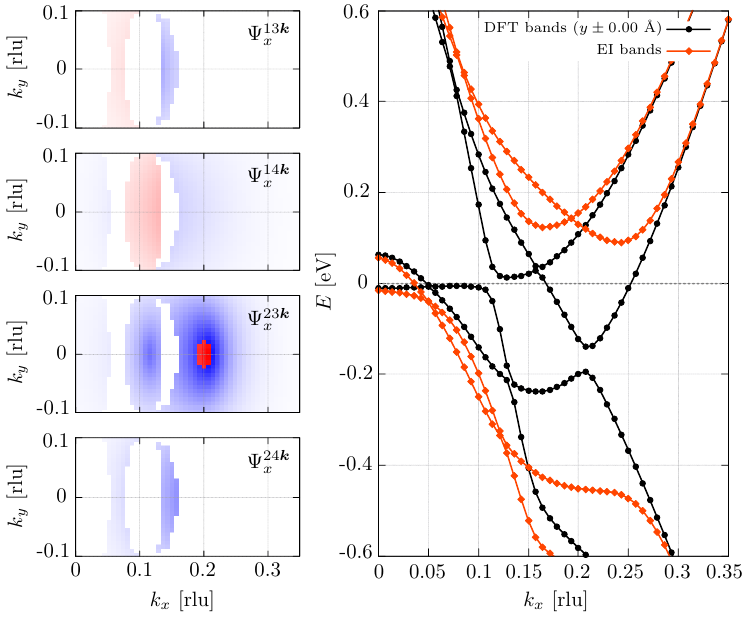}
\caption{
{\bf Bethe-Salpeter and EI gap equations within the four-band model.} 
Exciton wave function (left) and EI band reconstruction (right) within the four-band model for $\delta_{H_1}=0$ (GMS structure). 
The panels on the left show the exciton wave function components, $\Psi_x^{vc\bm{k}}$, where $v=1,2$ and $c=3,4$ are the indices of the two valence and two conduction bands, in ascending order of energy.
The range of the color scale is $[-0.1,0.1]$, with positive [negative] values in red [blue]. 
As in Fig.~3 of the main text, the maximum amplitude is reached at the two locations along the $\Gamma $X cut of the energy gaps separating bands $2$ and $3$. 
The corrections to the bands in the EI phase, given by $\Delta_{ij}(\bm{k})$ (see main text), increase with the magnitudes of the amplitudes $\Psi_x^{vc\bm{k}}$. 
Note that, with respect to Fig.~3 of the main text, here the Brillouin zone sampling is reduced along the $k_y$ direction. This significantly impacts the exciton binding energy, thus a lower screening ($\alpha_{\mathrm{2D}} = 1.9$ {\AA}) has been used to gap the EI band structure.
}
\label{fig:sifig01}
\end{figure} 

\subsection{Self-consistent mean-field theory of the excitonic insulator}

As for the two-band model illustrated in the main text, also in the present four-dimensional framework the EI ground state, $|\Psi_{\mathrm{EI}}\rangle$, is the Slater determinant of the filled, distorted valence band Bloch states,
\begin{equation}
    |\Psi_{\mathrm{EI}}\rangle =
    \prod_{i=1,2}\prod_{\bm{k}} \hat{\alpha}^{\dagger}_{i\bm{k}} 
    |\mathrm{vac}\rangle \, .
\end{equation}
Here $|\mathrm{vac}\rangle$ is the vacuum state with no electrons, $\hat{\alpha}^{\dagger}_{i\bm{k}}$ is the Fermi operator that creates an electron of momentum $\bm{k}$ in the $i$th renormalized band of the excitonic insulator,    
\begin{equation}\label{eq:alphadef}
    \hat{\alpha}_{i\bm{k}} = \sum_j a_{ij\bm{k}} \;\hat{a}_{j\bm{k}} \, ,
\end{equation}
$\hat{a}^{\dagger}_{j\bm{k}}$ creates an electron in the $j$th (conduction or valence) band of the pristine semimetal,
and we convene that $i=1,2$ and $i=3,4$ respectively label the EI valence and conduction bands.
The coefficients $a_{ij\bm{k}}$ occurring in \eqref{eq:alphadef} form a unitary matrix for given $\bm{k}$, which must be determined self-consistently.
The $a_{ij\bm{k}}$'s are the eigenvectors resulting from the diagonalization of the mean-field Hamiltonian, $\hat{H}_{\text{EI}}(\bm{k})$, the $4\times 4$ matrix
whose eigenvalues, $E_i(\bm{k})$, are the renormalized EI bands:
\begin{equation}
\label{eq:mat_4x4_Delta}
  H_{\text{EI}}(\bm{k}) =  \begin{pmatrix}
        \varepsilon_1(\bm{k}) & 0 & \Delta_{13}(\bm{k}) & \Delta_{14}(\bm{k}) \\
        0 & \varepsilon_2(\bm{k}) & \Delta_{23}(\bm{k}) & \Delta_{24}(\bm{k}) \\
        \Delta_{31}(\bm{k}) & \Delta_{32}(\bm{k}) & \varepsilon_3(\bm{k}) & 0 \\
        \Delta_{41}(\bm{k}) & \Delta_{42}(\bm{k}) & 0 & \varepsilon_4(\bm{k})
    \end{pmatrix} \, .
\end{equation}
Here the pristine bands, $\varepsilon_i(\bm{k})$, are the eigenvalues of the noninteracting Hamiltonian, $\hat{H}(\bm{k})=\hat{H}_0+\hat{H}_1$ [see \eqref{eq:H0} and 
\eqref{eq:H1}], and the (real) self-consistent interband hybridization,
$\Delta_{ij}(\bm{k}) = \Delta_{ji}(\bm{k})$, is driven
by the population of the exciton condensate. This term, which generalizes to multiple bands the gap function $\Delta(\bm{k})$ discussed in the main text, must be determined by solving the ``gap equation'', which complements the iterative cycle:
\begin{equation}
\label{eq:Dijk_SI}
   \Delta_{ij}(\bm{k}) =  
   \sum_{i',j',\bm{k}'} 
   W_{ij\bm{k}}^{i'j'\bm{k}'}
   \sum_{l,l'}\, 
   [a^{-1}_{\bm{k}'}]_{j' l} \,
   [a^{-1}_{\bm{k}'}]_{i' l'}\,
   \langle \hat{\alpha}^{\dagger}_{l\bm{k}'} \hat{\alpha}_{l'\bm{k}'} \rangle_{\Psi_{\mathrm{EI}}} \, .
\end{equation}
Note that in \eqref{eq:Dijk_SI}---and only there---the indices $i,i'=1,2$ and $j,j'=3,4$ run only over the valence and conduction states, respectively,
$W_{ij\bm{k}}^{i'j'\bm{k}'}$ is given by \eqref{eq:Wcvk_SI},
and the coefficients $[a^{-1}_{\bm{k}}]_{j l}$ are obtained by inverting
the eigenvector matrix $a_{ij\bm{k}}$. 
We evaluate the quantum (and thermal) average $\langle \hat{\alpha}^{\dagger}_{l\bm{k}} \hat{\alpha}_{l'\bm{k}} \rangle_{\Psi_{\mathrm{EI}}}$ occurring
in \eqref{eq:Dijk_SI} as
\begin{equation}\label{eq:mu}
\langle \hat{\alpha}^{\dagger}_{l\bm{k}'} \hat{\alpha}_{l'\bm{k}'}\rangle_{\Psi_{\mathrm{EI}}} = \delta_{l,l'}\, f[E_l(\bm{k})]\,,
\end{equation}
where $f(\cdot)$ is the Fermi-Dirac distribution function. 
At zero temperature, for the intrinsic excitonic insulator phase, 
we set $f$ to $1$ for valence states and $0$ for conduction states. 

The numerical self-consistent cycle starts by using the components of the lowest-energy exciton, $\Psi_{x}^{ij\bm{k}}$, derived from the solution of the four-band Bethe-Salpeter equation \eqref{eq:BSE_SI}, as a seed for
$ \Delta_{ij}(\bm{k})$ that enters the matrix \eqref{eq:mat_4x4_Delta}. 
After $\hat{H}_{\text{EI}}$ is diagonalized, the resulting orthogonal eigenvector matrix is inverted to evaluate the coefficients $[a^{-1}_{\bm{k}}]_{j l}$ entering \eqref{eq:Dijk_SI}.
The new value of $ \Delta_{ij}(\bm{k})$ is then inferred by performing the sum over $\bm{k}'$ in \eqref{eq:Dijk_SI}, and then used to update \eqref{eq:mat_4x4_Delta} at the next step. 
The cycle is repeated until $ \Delta_{ij}(\bm{k})$ converges for all $\bm{k}$, $i$, $j$. 

For completeness, we note that, in the presence of doping, or at finite temperature, or in the `excitonic semimetal' scenario, the set of self-consistent equations reported above must be complemented by an additional condition, the total charge balance, which implicitly defines the chemical potential, $\mu$. Such equation must be solved for $\mu$ at each step of the 
iterative cycle, as the chemical potential enters the average \eqref{eq:mu} through the Fermi distribution function. In principle, also the screened interaction $W$ should be evaluated self-consistently, as the variation of 
$\mu$ impacts the screening of charge carriers \cite{Kozlov1965}. In the simplest case of an intrinsic excitonic insulator at zero temperature, $\mu$ lies in the gap and the charge balance condition is unnecessary.  

The method is illustrated in the right column of
Supplementary Fig.~\ref{fig:sifig01}, which compares the noninteracting 
and EI band structure 
for $\delta_{H_1}=0$ and $\alpha_{\mathrm{2D}}=1.9$ {\AA}, with a $50\times29$ $\bm{k}$-point grid. Note that the renormalization of the bottom conduction and top valence bands due to e-h interaction compares with that assessed within the two-band model, as shown in Fig.~3 of main text.

\subsection{Discussion}

The four-band model reported in this Supplementary Note provides valuable insight into the understanding of the variation of the band structure with $y$, the nature of the condensing exciton, and the consequent spontaneous breaking of the crystal symmetry. 
On the other hand, it has a few caveats, which we discuss in the following and that ultimately led us to present the two-band results in the main text.

\begin{figure}[h]
\includegraphics[width=0.55\linewidth]{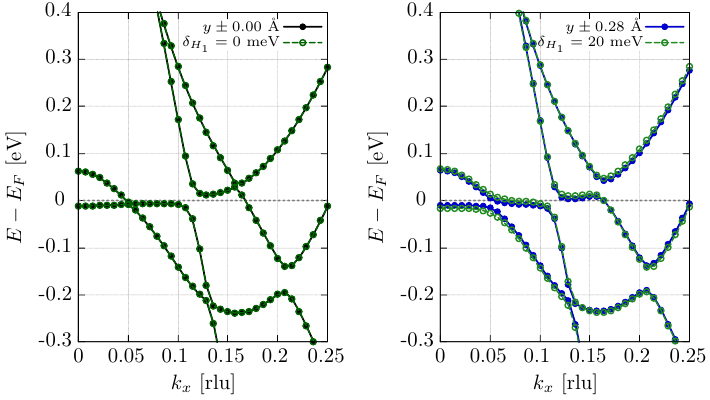}
\caption{
{\bf Comparison between model and first-principles bands.}
Left: Model noninteracting bands, computed for $\delta_{H_1}=0$ (open circles), compared to first-principles bands for the GMS structure, $y=0$ (filled circles).
Right: Model noninteracting bands, computed for $\delta_{H_1}=20$ meV (open circles), compared to first-principles bands of the bilayer structure with the top layer displaced by $\pm0.28$ {\AA} (filled circles). 
In both panels the bands are plotted along the $\Gamma$X cut of Brillouin zone ($k_y = 0$).
}
\label{fig:siaddedfig02}
\end{figure}

The first issue is the modeling of layer sliding through the envelope-function Hamiltonian $\hat{H}_1$ of \eqref{eq:H1}, which does not depend on $\bm{k}$.
This simplification allows for assessing the ``frozen phonon'' parameter, $\delta_{H_1}$, by maximizing the overlap with the target DFT bands computed for a given sliding displacement. For example, for $\pm 0.280$ {\AA} displacement of the top layer from the GMS structure, we find $\delta_{H_1} = 20$ meV (right panel of Supplementary Figure \ref{fig:siaddedfig02}).
However, a non-local, $\bm{k}$-dependent $\delta_{H_1}$ parameter would be needed to match exactly the evolution of band anticrossing with $y$, as predicted from density functional theory. This matter is illustrated in Supplementary Fig.~\ref{fig:siaddedfig02}. 
Whereas in the left panel the matching of first-principles (filled circles) and model (open circles) bands is complete by construction ($y=0$), in the right panel ($y=\pm0.28$ {\AA}) the model conduction band crossing the Fermi energy around $k_x=(0.15)$$\,2\pi/a$ overestimates the gap opening, while the valence band underestimates it. 
Though the discrepancy is tiny on the energy scale of the plot, the total energy gain associated with exciton binding is amplified by condensation, leading to a strong renormalization of the sliding barrier, as discussed in the main text. 

Another concern is the envelope-function approximation at the basis of the four-band model, whose reliability deteriorates as one moves from the Brillouin zone center. 
The issue becomes severe for $\left|k_y\right|  >  (0.1)\,2\pi/b$, due to the anticrossing behavior of bands that is not well reproduced by the model. 
This limits our ability to check the convergence of the solution of the Bethe-Salpeter equation \eqref{eq:BSE_SI} for large interaction strength. 
For this reason we relied eventually on the two-band model of the main text, based on energy bands purely derived from first principles, which allows us to extend arbitrarily the Brillouin zone sampling.

%%%%%%%%%%%%%%%%%%%%%%%%%%%%%%%%%%%%%%%%%%%%%%
%%%%%%%%%%%%%%% PLOTS VS a2D %%%%%%%%%%%%%%%%%
\clearpage
\section{Sensitivity of model to parameters and validation from first principles}

Bilayer WTe$_2$ exhibits a tiny gap of a few meV, excitonic in nature \cite{sun_evidence_2022}, but is predicted to be a semimetal from DFT (Fig.~2 of main text). 
This poses a formidable computational challenge, since a full first-principles treatment of excitons would require the calculation of the GW correction to band structure, which includes dynamical screening effects, as well as the evaluation of the intraband contribution to electronic polarizability. 
Such high-level tasks are essential to capture excitonic effects in the presence of a two-dimensional Fermi surface \cite{Liang2015, leon_frequency_2021, champagne_quasiparticle_2023, leon_efficient_2023, guandalini_efficient_2024, sesti_efficient_2025}. 
Furthermore, the calculation of the EI phase is inherently self-consistent, in principle including the re-evaluation of screening at each step of the iterative cycle. 
This is relevant, as the static and dynamical GW correlations tend to decrease the semimetal band overlap and increase the range of the e-h attraction that binds excitons, which in turn may condense and open the gap. 
Eventually, the screening behavior of the reconstructed EI ground state is compatible with that of a narrow-gap semiconductor rather than that of the initial semimetal state.

Since the GW description of the bilayer is presently out of reach, we developed the two-band model approach of the main text. 
Luckily, two first-principles investigations recently reported consistent results for monolayer WTe$_2$ \cite{sun_evidence_2022,wu_quasiparticle_2024}, which we can use to validate the bilayer model, as we expect similar screening behaviour and exciton binding. 
Namely, Wu {\it et al.}, starting from a DFT semimetal ground state, showed \cite{wu_quasiparticle_2024} the tendency of the GW self-energy (in both flavors $G_0W_0$ and self-consistent COHSEX) to remove the Fermi surface by opening a gap. 
We expect a similar effect for the bilayer, which will tend to suppress the intraband contribution to screening. Importantly, the GW-corrected band structure by Wu {\it et al.} compares well with the one we derived previously (via hybrid DFT) to compute excitons from first principles \cite{sun_evidence_2022}. 
Therefore, below we use the findings of Ref.~\onlinecite{sun_evidence_2022} for the monolayer to benchmark those of the bilayer.

The two-band model critically relies on the parametrization of the screened e-h Coulomb attraction $W$ via the polarizability $\alpha_{\mathrm{2D}}$ [equation (3) of main text]. 
We cannot fix $\alpha_{\mathrm{2D}}$ from the DFT calculation, since the predicted ground state is semimetallic: The occurrence of the spurious Fermi surface implies an intraband contribution to $\alpha_{\mathrm{2D}}$, which diverges at long wave length. 
The only consistent way to fix $\alpha_{\mathrm{2D}}$ is to choose a value that results in the opening of a gap, which hence suppresses the intraband polarizability. 
Indeed, the chosen value of $\alpha_{\mathrm{2D}}=4.3$ {\AA} provides a tiny many-body indirect gap that compares with experiments \cite{sun_evidence_2022}. 
The resulting gap is 3 meV for $y=0$ and 7 meV for $y=\bar{d}$ (Fig.~3b of main text, EI band structure).

\begin{figure}[h]
\includegraphics[width=0.35\linewidth]{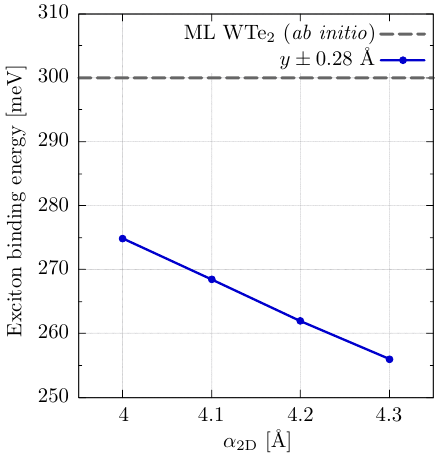}
\caption{
{\bf Exciton binding energy upon variation of the screening parameter.}
Binding energy of the lowest exciton, resulting from the solution of Bethe-Salpeter equation (1) of main text, versus screening parameter $\alpha_{\mathrm{2D}}$. 
The results were obtained for sliding displacement $y \pm 0.28$ {\AA}, using a $30\times60$ $\bm{k}$-point grid sampling the region $k_x  \in [0,0.3]$ rlu and $k_y \in [-0.3,0.3]$ rlu. 
The lines connecting the dots are a guide to the eye. 
The dashed line highlights the exciton binding energy for monolayer (ML) WTe$_2$, computed from first principles \cite{sun_evidence_2022}.
}
\label{fig:siaddedfig03}
\end{figure}

We may justify further the chosen value of $\alpha_{\mathrm{2D}}$ by contrasting the exciton binding energy inferred from $W$ with the first-principle prediction for the monolayer \cite{sun_evidence_2022}. 
This is illustrated in Supplementary Fig.~\ref{fig:siaddedfig03}, which compares the model exciton binding energy (dots) with the reference value \cite{sun_evidence_2022} of 300 meV for the monolayer (dashed line), in the range $4\leq\alpha_{\mathrm{2D}}\leq4.3$ {\AA}.  
In general, as the screening of the e-h attractive potential increases with $\alpha_{\mathrm{2D}}$, the exciton binding energy decreases. 
Importantly, in the plotted range of $\alpha_{\mathrm{2D}}$ the binding energy for the bilayer is smaller than the one for the monolayer, which matches the physical expectation of stronger screening in the bilayer.
We disregard the outer range of $\alpha_{\mathrm{2D}}$, since for $\alpha_{\mathrm{2D}}\ll4$ {\AA} the bilayer binding energy nonphysically exceeds the monolayer value, whereas for $\alpha_{\mathrm{2D}}>4.3$ {\AA} the resulting excitonic phase is semimetallic and the Rytova-Keldish form of $W$ loses accuracy. 

\begin{figure}[h]
\includegraphics[width=0.6\linewidth]{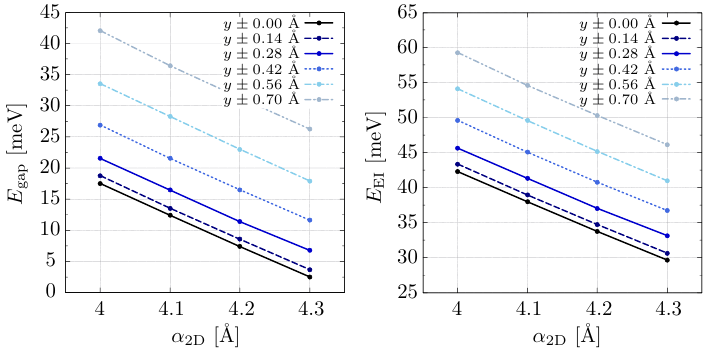}
\caption{
{\bf Energetics of the excitonic insulator vs screening and sliding displacement.}
Many-body indirect gap in the EI phase, $E_{\mathrm{gap}}$ (left panel), and total energy gain, $E_{\mathrm{EI}}$ (right panel), versus screening parameter, $\alpha_{\mathrm{2D}}$, for various values of the sliding displacement, $y$. 
 The connecting lines are a guide to the eye. 
}
\label{fig:sifig02}
\end{figure}

Like the exciton binding energy, also the EI gap function, $\Delta(\bm{k})$, converges to smaller values with increasing $\alpha_{\mathrm{2D}}$, which translates into weaker renormalization of EI bands and smaller size of the many-body indirect gap, $E_{\mathrm{gap}}$.
Remarkably, whereas $E_{\mathrm{gap}}$ is significantly sensitive to $\alpha_{\mathrm{2D}}$, the height of the sliding barrier potential is not. 
The reason is that the excitonic renormalization of the barrier height depends on the \emph{variation} of $E_{\mathrm{gap}}$ with $y$, and not on its absolute magnitude. 
This is illustrated by Supplementary Fig.~\ref{fig:sifig02}, which plots both $E_{\mathrm{gap}}$ (left panel) and the EI total energy gain, $E_{\mathrm{EI}}$ (right panel), vs $\alpha_{\mathrm{2D}}$, for different sliding displacements, $y$. 
Importantly, both curves are essentially parallel straight lines, namely, the variation of both $E_{\mathrm{gap}}$ and $E_{\mathrm{EI}}$ with $y$ does not depend on the screening parameter, $\alpha_{\mathrm{2D}}$. 
In conclusion, the key result of our work, the excitonic renormalization of the height of the sliding barrier, is robust against the details of the two-band model.

%%%%%%%%%%%%%%%%%%%%%%%%%%%%%%%%%%%%%%%%%%%%%%
%%%%%%%%%%%%%%% SOC vs noSOC %%%%%%%%%%%%%%%%%
\clearpage
\section{Spin-orbit coupling}

\begin{figure}[h]
\includegraphics[width=0.6\linewidth]{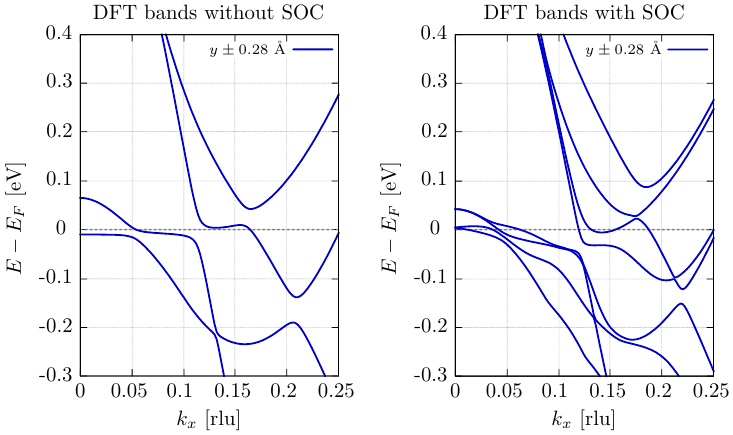}
\caption{
{\bf Effect of spin-orbit coupling on DFT bands.}
DFT-PBE band structure without (left panel) and with (right panel) spin-orbit coupling (SOC). Results for the bilayer structure with sliding displacement $y \pm 0.28$ {\AA}. The parameters of the DFT calculation with SOC are the same as in the Methods section of main text, with the exception that the corresponding full-relativistic pseudopotentials from the same library have been used \cite{hamann_optimized_2013}.
}
\label{fig:siaddedfig04}
\end{figure}

In the natural stacking of bilayer WTe$_2$ the inversion symmetry of the single-layer crystal structure is not preserved. 
As a result, spn-orbit coupling (SOC) splits the bands of bilayer WTe$_2$ by breaking Kramers degeneracy
(Supplementary Fig.~\ref{fig:siaddedfig04}), while Bloch states remain doubly degenerate in monolayer WTe$_2$ \cite{sun_evidence_2022}. 
Thus, the splitting effect must be attributed to inter-layer interaction, and may be tentatively rationalized as a Rashba-like term, $\sim \bm{S\cdot(k\times E)}$, with $\bm{S}$ being the spin and $\bm{E}$ the microscopic electric field, after noting that the band splitting vanishes as $\bm{k}\rightarrow 0$ (right panel of Supplementary Fig.~\ref{fig:siaddedfig04}). 

Since our key findings depend on the variation of the total energy gain of the EI phase, $E_{\mathrm{EI}}$, with the sliding displacement $y$, we expect the inclusion of SOC not to alter qualitatively our conclusions. 
In fact, even if the energy bands split with the SOC field, we expect the Rashba-like corrections, $\sim\pm \left|\bm{k\times E}\right|$, to substantially cancel out when summing over occupied states to compute $E_{\mathrm{EI}}$. 
An accurate statement would require to work out in detail the full numerical inclusion of SOC effects, as the intricacies of anticrossings shown in right panel of Supplementary Fig.~\ref{fig:siaddedfig04} 
may compromise the reliability of any modelization. 
Such a task is beyond the scope of the present work.

On the other hand, SOC will qualitatively change the microscopic symmetries spontaneously broken by exciton condensation, as already mentioned in Supplementary Section \ref{s:broken}. 
In particular, a magnetic instability could add to the breaking of the mirror symmetry $x\rightarrow -x$. 
The inclusion of SOC effects would significantly complicate the theory of the EI phase, by increasing the number of independent gap equations to solve self-consistently.

%%%%%%%%%%%%%%%%%%%%%%%%%%%%%%%%%%%%%%%%%%%%%%
%%%%%%%%%%%%%%% BIBLIOGRAPHY %%%%%%%%%%%%%%%%%
\clearpage
\twocolumngrid
\bibliography{references.bib}

\end{document}